\documentclass[sigconf]{acmart}
\pdfoutput=1
\AtBeginDocument{%
  \providecommand\BibTeX{{%
    \normalfont B\kern-0.5em{\scshape i\kern-0.25em b}\kern-0.8em\TeX}}}

\setcopyright{acmcopyright}
\copyrightyear{2021}
\acmYear{2021}
\acmDOI{TBD/TBD}

\acmConference[TBD]{TBD}{TBD}
\usepackage{amsmath}
\usepackage{multicol}

\usepackage{outlines}
\usepackage{xcolor}
\definecolor{keywordcolor}{rgb}{0.941, 0.875, 0.686}
\definecolor{emphcolor}{rgb}{0.486, 0.722, 0.733}
\definecolor{commentcolor}{rgb}{0.498, 0.624, 0.498}
\definecolor{backgroundcolor}{rgb}{0.247, 0.247, 0.247}
\definecolor{stringcolor}{rgb}{0.8, 0.576, 0.576}
\definecolor{basiccolor}{rgb}{0.863, 0.859, 0.796}
\definecolor{numbercolor}{rgb}{0.310, 0.310, 0.310}
\definecolor{LightGray}{rgb}{0.898, 0.898, 0.898}
\definecolor{LighterGray}{rgb}{0.95294, 0.95294, 0.95294}

\usepackage{caption}
\usepackage[section,frozencache=true,cachedir=.]{minted} 
\newenvironment{code}{\captionsetup{type=listing}}{}

\usepackage[many]{tcolorbox}
\tcbuselibrary{minted}
\newtcbinputlisting{\mycode}[2]{%
  listing engine=minted,
  minted language={#1},
  listing file={#2},
  minted options={
    xleftmargin=2em,
    autogobble=true,
    numbers=left,
    numbersep=2em,
    baselinestretch=1.0,
    fontsize=\footnotesize,
    breaklines=true
    },
  listing only,
  breakable,
  enhanced jigsaw,
  colframe=black,
  sharp corners,
  boxrule=1pt,
  colback=lightgray!10,
  left=-1.5em,
  left skip=2em,
  width=\linewidth-2em
}

\usepackage{microtype}

\newcommand{\eq}[1]{Eq.~(\ref{eq:#1})}
\newcommand{\fig}[1]{Fig.~\ref{fig:#1}}

\newcommand{\sect}[1]{Sec.~\ref{sec:#1}}

\newcommand{\lst}[1]{Lis.~\ref{lst:#1}}
\newcommand{\seconds}{{~}\textrm{s}}
\newcommand{\femtoseconds}{{~}\textrm{fs}}
\newcommand{\micrometers}{{~}\mu\textrm{m}}
\newcommand{\nanometers}{{~}\textrm{nm}}
\newcommand{\comp}{{~}c/\omega_{\rm pe}}
\newcommand{\omp}{{~}\omega_{\rm pe}}
\newcommand{\nenaut}{n_{\rm e0}}
\newcommand{\ninaut}{n_{\rm i0}}
\newcommand{\nnaut}{n_{0}}
\newcommand{\me}{m_{\rm e}}

\newcommand{\nppc}{n_{\rm ppc0}}
\newcommand{\helvet}[1]{{\fontfamily{phv}\selectfont #1}}

\begin{document}

\title{In-Situ Assessment of Device-Side Compute Work for Dynamic Load Balancing in a GPU-Accelerated PIC Code}

\author{Michael E. Rowan}
\email{mrowan@lbl.gov}
\orcid{0000-0003-2406-1273}
\author{Kevin N. Gott}
\email{kngott@lbl.gott}
\orcid{0000-0003-3244-5525}
\author{Jack Deslippe}
\email{jrdeslippe@lbl.gov}
\orcid{0000-0003-1785-4187}
\affiliation{%
  \institution{Lawrence Berkeley National Laboratory, NERSC}
  \streetaddress{1 Cyclotron Road, Building 59}
  \city{Berkeley}
  \state{California}
  \country{USA}
  \postcode{94720}
}
\author{Axel Huebl}
\email{axelhuebl@lbl.gov}
\orcid{0000-0003-1943-7141}
\author{Maxence Th{\'e}venet}
\authornote{Now with Deutsches Elektronen Synchrotron (DESY), Germany}
\email{maxence.thevenet@desy.de}
\orcid{0000-0001-7216-2277}
\author{Remi Lehe}
\email{rlehe@lbl.gov}
\orcid{0000-0002-3656-9659}
\author{Jean-Luc Vay}
\email{jlvay@lbl.gov}
\orcid{0000-0002-0040-799X}
\affiliation{%
  \institution{Lawrence Berkeley National Laboratory, ATAP}
  \streetaddress{1 Cyclotron Road, Building 71}
  \city{Berkeley}
  \state{California}
  \country{USA}
  \postcode{94720}
}
\renewcommand{\shortauthors}{M. E. Rowan, A. Huebl, K. N. Gott, J. Deslippe, M. Th\'{e}venet, R. Lehe, and J.-L. Vay}

\begin{abstract}
Maintaining computational load balance is important to the performant behavior of codes which operate under a distributed computing model. This is especially true for GPU architectures, which can suffer from memory oversubscription if improperly load balanced. We present enhancements to traditional load balancing approaches and explicitly target GPU architectures, exploring the resulting performance. A key component of our enhancements is the introduction of several GPU-amenable strategies for assessing compute work. These strategies are implemented and benchmarked to find the most optimal data collection methodology for in-situ assessment of GPU compute work.  For the fully kinetic particle-in-cell code \helvet{WarpX}, which supports MPI+CUDA parallelism, we investigate the performance of the improved dynamic load balancing via a strong scaling-based performance model and show that, for a laser-ion acceleration test problem run with up to 6144 GPUs on Summit, the enhanced dynamic load balancing achieves from $62\%$--$74\%$ ($88\%$ when running on 6 GPUs) of the theoretically predicted maximum speedup; for the 96-GPU case, we find that dynamic load balancing improves performance relative to baselines without load balancing ($3.8{\times}$ speedup) and with static load balancing ($1.2{\times}$ speedup). Our results provide important insights into dynamic load balancing and performance assessment, and are particularly relevant in the context of distributed memory applications ran on GPUs.
\end{abstract}



\settopmatter{printfolios=true} 
\maketitle

\section{Introduction}
\label{sec:intro}
GPU-accelerated machines entered the TOP500 rankings just over a decade ago \cite{top500}. 
GPUs offer an effective path to high computational throughput, and GPU accelerated machines are particularly of interest, as they are central to the US Exascale, and EuroHPC Pre-Exascale initiatives.
HPC software must be adapted to take full advantage of heterogeneous systems as parallelism continues to increase and we approach an era of exascale computing in leadership-class supercomputers.
Maintaining equal distribution of computational load (\textit{dynamic load balancing}) is crucial for distributed memory applications to make efficient use of hardware with continually increasing parallel-compute capabilities \cite{Germaschewski2016, Tsuzuki2016, Miller2020}.

In this paper, we present and investigate GPU-targeted enhancements for in-situ dynamic load balancing. These techniques are demonstrated in the particle-in-cell (PIC) framework \helvet{WarpX} \cite{Vay2018}, which, as a particle-mesh simulation code, models fields via an Eulerian description and particles via a discretization in Lagrangian markers.
In \helvet{WarpX}, multi-node parallelism is achieved through spatial domain decomposition, which is supported through the block-structured framework \helvet{AMReX} \cite{Zhang2019}.
\helvet{WarpX}'s particles are used to model kinetic phenomena and often cause localized spikes in memory consumption and compute demand.
It is thus likely, that herein studied scenarios exemplifying realistic geometries of contemporary research, using highly mobile particles are applicable to other distributed memory codes.
In particular, \helvet{WarpX} models plasma physics problems, including laser-plasma acceleration and particle acceleration in astrophysical contexts \cite{Vay2018}.

The outline of this paper is as follows: 
In \sect{lb_strategy}, we introduce load balancing essentials for particle-mesh codes and present our dynamic load balancing improvements for GPU systems.
In \sect{test_problem}, we introduce laser-ion acceleration as a scientifically relevant test problem with which to explore the performance of our improvements to \helvet{WarpX}'s dynamic load balancing. 
Next, in \sect{performance}, we discuss the performance of \helvet{WarpX}'s enhanced dynamic load balancing in the context of a strong scaling-based performance model, and present weak scaling tests of dynamic load balancing performance.
We conclude in \sect{conclusion}, with a summary of the generality of the proposed approach, advantages relative to related works, and discussion of future work.

\section{In-Situ Device-Side Dynamic Load Balancing}
\label{sec:lb_strategy}
In this section, we describe novel methods we have devised for measuring device-side compute work at runtime and the dynamic load balancing algorithms used with these measurements.  

In \sect{amrex_overview}, we introduce general features of distributed memory particle-mesh codes that are necessary for a discussion of load balancing and introduce the terminology used in the software framework \helvet{AMReX}.
In \sect{lb_warpx}, we present details of our enhancements to dynamic load balancing when using GPUs, with a focus on the challenges of load balancing in a GPU-based code and the GPU-specific aspects of the load balance algorithm. 

\subsection{Overview of the Domain Decomposition in Particle-Mesh Codes}
\label{sec:amrex_overview}
\begin{figure}[!ht]
  \centering
  \includegraphics[width=\linewidth, trim={0.0cm 0.0cm 0.0cm 0.0cm}]{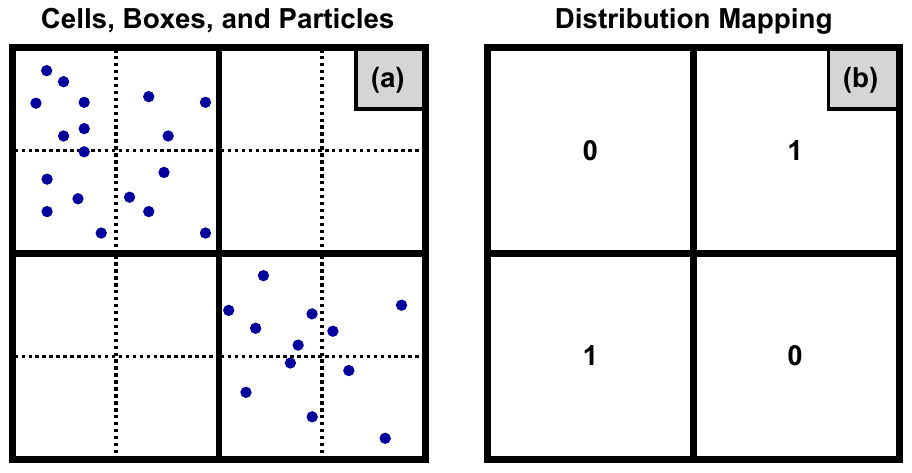}
  \caption{Example domain with 16 cells (delimited by dotted lines) and 4 sub-domains (delimited by solid lines; each sub-domain contains 4 cells) (a); blue circles represent particles, and are associated with the sub-domain with which they overlap. A distribution mapping indicates which MPI rank manages each box (b); this example demonstrates load imbalance because rank~0 manages 30 particles, whereas rank~1 manages no particles.}
  \label{fig:AMReXOverview}
\end{figure}

A common motif in parallel particle-mesh codes is to partition the simulation domain into separate sub-domains, typically contiguous spatial sub-domains. These sub-domains are interchanged between compute elements at runtime to achieve an even distribution of compute work across computational resources. A load-balancing algorithm is implemented to achieve this even distribution. 
An example domain consisting of 16 (rectilinear) cells decomposed into 4 sub-domains (delimited by solid lines; each sub-domain contains 4 cells) is shown in panel (a) of \fig{AMReXOverview}. A sub-domain usually contains the field data that describes its contiguous space, as well as associated particle data for the particles that lie within the bounds of that sub-domain.
For the example in panel (a) of \fig{AMReXOverview}, the boxes \\ in the upper left and lower right have 18 and 12 associated particles, respectively. 

Within the MPI parallelism model, which is ubiquitous in distributed memory codes, it is natural to assign each sub-domain to one exclusive MPI rank; for MPI+$X$ parallelism (where $X$ corresponds to an accelerator platform, e.g., CUDA), the ranks correspond to unique GPUs.

Our study is designed as a general improvement in \helvet{AMReX}, a performance-portable, block-structured framework which provides infrastructure for mesh-based simulations, with capability to support particles \cite{Zhang2019}. 
For the remainder of this paper, we will use \helvet{AMReX} terminology, as it is consistent with our study and the naming convention is highly descriptive.
In \helvet{AMReX}, a `sub-domain' is a rectilinear region of cells, defined by a \textit{box}. A \textit{distribution mapping} describes the current MPI process ownership of field and particle data associated with each box; since in our study we use one MPI rank per GPU, the distribution mapping also maps the GPU ownership of each box. This consists of a box-sized vector containing the MPI rank (equivalently, GPU ID) to which each box is assigned.
For the example in panel (a) of \fig{AMReXOverview}, one possible distribution mapping is shown in panel (b); the boxes (including both their cells and particles) in the upper left and lower right belong to rank~0 (also GPU 0), and the boxes in the upper right and lower left belong to rank~1 (also GPU 1).

Our load-balancing strategy is tested in \helvet{WarpX}, a particle-mesh application, implemented on top of \helvet{AMReX}.
Cells are updated in self-consistent, explicit time steps solving Maxwell's equations.
Throughout all cells, representative particle markers (weighted particles) move according to the Lorentz-force and in turn modify fields through deposited, charged currents \cite{Birdsall1985}.
When ran on systems with accelerators, locally assigned boxes of particles and structured field (cell) data are stored persistently in device memory.
Implementing advanced, numerical schemes for those operations, single-source kernels consistently accelerate the whole application.

Load balancing in \helvet{WarpX} consists of a global update of the distribution mapping across all MPI ranks, and redistribution of particle and cell data in accordance with the new distribution mapping (this is further discussed in the next section, \sect{lb_warpx}).
The example configuration in \fig{AMReXOverview} demonstrates particle load imbalance; while ranks 0 and 1 each manage 8 cells, rank 0 manages 30 particles and rank 1 manages no particles.

\subsection{Load Balancing Strategy}
\label{sec:lb_warpx}
A basic outline for dynamic load balancing in a distributed memory particle-mesh code (representative of what we use in the present work) is sketched in \lst{lb_strategy}; this resides within the main time-stepping loop in \helvet{WarpX}.

The frequency of the load balancing calculation is controlled by a user-selected load balance interval (line 1).  
Computational costs corresponding to each box are gathered from all GPUs to the root process, so that the root process has a global view of the current GPU ownership and cost for each box over the full simulation domain. With this information, the root process computes a load balanced distribution mapping and the corresponding \textit{load balance efficiency} $E$ for the old and new mappings (lines 9--11). \\%
\begin{code}
\caption{Dynamic load balancing routine\label{lst:lb_strategy}}%
\mycode{c++}{figures/loadBalanceRoutine.cpp}
\end{code}%
\noindent This load balance efficiency is defined as: 
\begin{align} \label{eq:lb_efficiency}
E &\equiv c_{\rm avg}/c_{\rm max}
\end{align}
where $c_{\rm avg}$ and $c_{\rm max}$ are the average and maximum costs, respectively, over all GPUs. The cost per GPU is taken as the sum of costs over all boxes managed by it; an efficiency of 1 indicates a perfectly balanced distribution. By default, the new distribution mapping will not be broadcast to all GPUs (line 17), but if the proposed efficiency exceeds the current efficiency by a user-selected amount (lines 18--21), the new distribution is communicated to all GPUs and updated (lines 23--31).
Update of the distribution mapping includes shuffling ownership of boxes as well as redistributing particles to GPUs (line 30), and when ran is the most expensive step of the load balancing routine (this is the case for simulations we present in \sect{parameter_dependence}; redistribution of box and particle data typically constitutes ${\gtrsim}99.7\%$ of the time required to load balance). Therefore, only redistributing in cases that will yield a substantial improvement is critical to achieving an optimal load balancing implementation.
We control this with a hyperparameter for the required improvement to the load balance efficiency $E$, i.e. a value that must be met or exceeded in order for the proposed distribution mapping to be communicated and updated (the performance impact of this hyperparameter is discussed in \sect{parameter_dependence}).

In distributing costs equally over all GPUs (i.e., computing a new distribution mapping; \lst{lb_strategy}, lines 9--11), a variety of strategies have been developed \cite{Bader2013,Germaschewski2016,Tsuzuki2016,Miller2020}.  
Two of the most common algorithms are \textit{knapsack}, which assigns sub-domains to GPUs such that the corresponding computational costs are spread as equally as possible over all GPUs, and \textit{space-filling curve}, in which sub-domains are enumerated with a Morton Z-order space-filling curve and the resultant ordering is partitioned so that costs are distributed as equally as possible among GPUs.
For our simulations of laser-ion acceleration, the performance impact of these algorithms when running on GPUs is discussed in \sect{parameter_dependence}.

\begin{figure}[!ht]
  \centering
  \includegraphics[width=\linewidth, trim={0.2cm 0.2cm -0.2cm 0.2cm}]{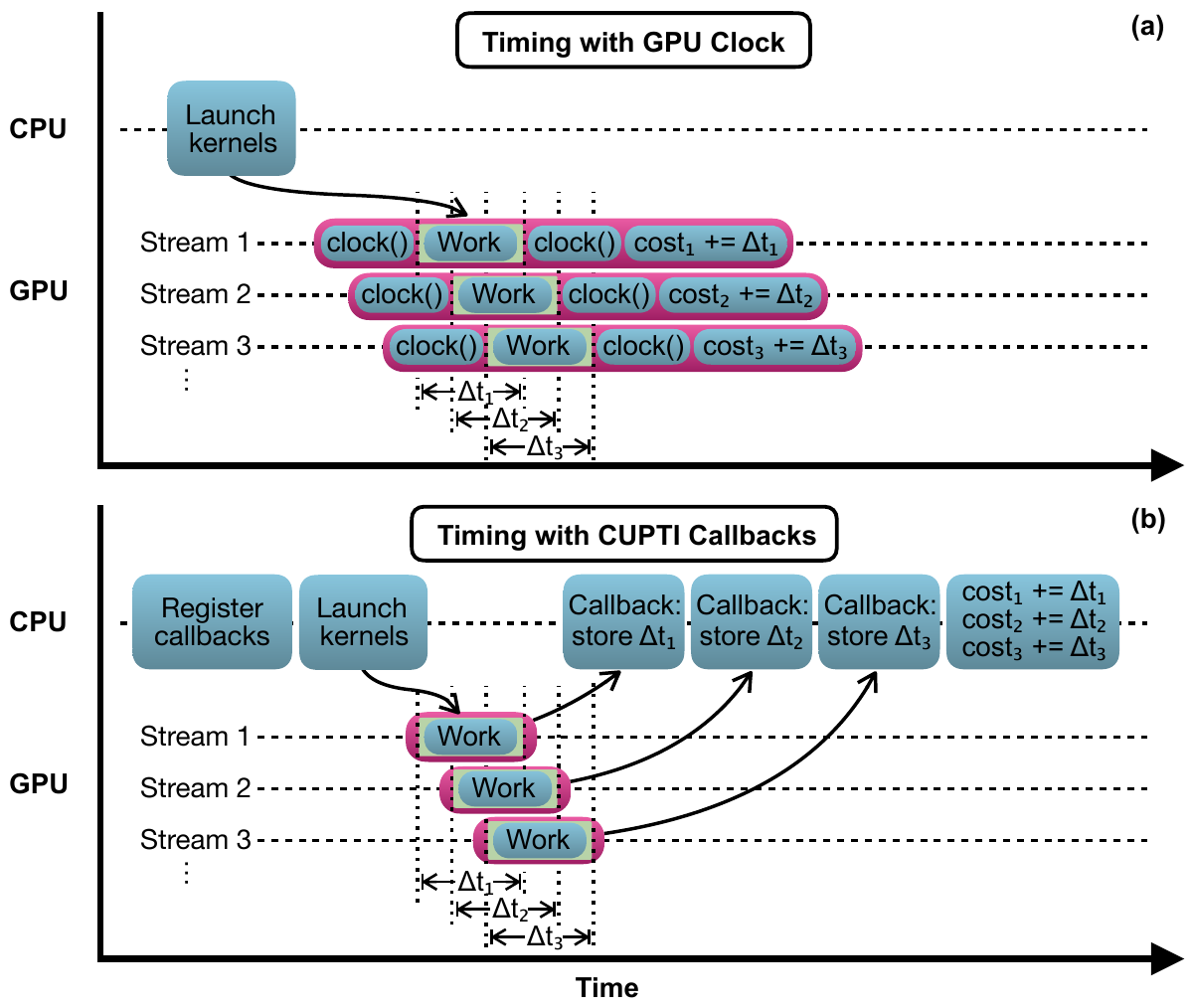}
  \caption{Schematic summary of GPU clock (a) and CUPTI (b) methods for assessing GPU compute work. 
  The diagram in (a) is only representative, in that the actual measured quantity in our implementation is summed thread execution time, rather than kernel execution time; with CUPTI (b), we do measure kernel times as a proxy for GPU compute.}
  \label{fig:KernelTiming}
\end{figure}

Measuring computational cost offers a unique challenge with GPUs, as GPUs leverage asynchronous compute \cite{Malony2011}.  
For this reason, timing sections of code with CPU timers will not yield a useful load balancing metric \cite{Dietrich2012}.
To address this challenge, we implemented three different GPU-amenable cost measurement strategies to estimate the compute work associated with a box, (\textit{Heuristic}, \textit{GPU clock}, and \textit{CUPTI}), which we summarize below:
\begin{itemize}
    \item Heuristic: Compute work is estimated as a weighted linear sum of the number of particles and cells per box.  
    Note that the optimal choice of weights may vary depending on hardware, choice of field solver, and the interpolation order of particle shapes.
    \item GPU clock: The device \texttt{clock()} function is used to measure thread execution time, which is accumulated using GPU atomic add operations; the procedure is shown schematically in \fig{KernelTiming}~(a). 
    To mitigate latency, thread times are accumulated in shared memory before their sum is transferred to global memory.
    With this GPU clock-based timer, we measure the thread-summed execution time of a compute-intensive kernel, which serves as a proxy for the compute work associated with a box. 
    For the laser-ion acceleration problem discussed in \sect{test_problem}, current deposition typically accounts for $50\%$ of the total walltime, with the remaining walltime dominated by communication routines, thus we take the time spent in current deposition as representative of the compute work that should be considered during load balance.
    \item CUPTI: With the CUDA Profiling Tools Interface (CUPTI) API \cite{CUPTIAPI}, we constructed a timer capable of accessing kernel execution times on-the-fly; the CUPTI-based timing strategy is summarized in \fig{KernelTiming} (panel (b)). 
    CUPTI enables collection of \textit{kernel activity records}--- data structures that contain kernel information including the absolute start and end times of the kernel. 
    Registered callback functions, which handle the request and delivery of buffers used to store activity records, are activated by GPU activity;
    when the buffer return is complete, the activity records stored therein can be used to compute kernel duration. 
    With this CUPTI-based kernel timer, we measure the duration of the current deposition kernel and use it as a proxy for the computational cost associated with a box.
\end{itemize}
The heuristic method has been used frequently in particle-mesh codes which run on CPU \cite{Zeyao2001, OSIRIS2002, Spitkovsky2005, VPIC2008, Pearce2014, Smilei2018, Miller2020}, and transfers readily to particle-mesh codes on GPU \cite{PIConGPU2013, Tsuzuki2016, Germaschewski2016}, given the user has tuned appropriately the particle and cell weights.  However, as mentioned above, the optimal choice of weights can vary depending on algorithmic choices and hardware, which limits the general applicability of heuristic cost measurement. 
The need for careful tuning by the user can be eliminated with an on-the-fly measurement technique such as GPU clock or CUPTI mentioned above; to the authors' knowledge, the present work represents the first implementation and application of such techniques for dynamic load balancing in a particle-mesh code running at scale on GPUs.


%
%
%

\section{Dynamic Load Balancing Test Problem: 2D Laser-Ion Acceleration}
\label{sec:test_problem}
In this section, we discuss the performance of \helvet{WarpX}'s improved dynamic load balancing in simulations of laser-ion acceleration physics.  
In \sect{setup}, we describe the problem setup and initial conditions of our simulations.
In \sect{time_evolution}, we show the time evolution of load balancing in our test problem (i.e., the time evolution of cost per computational grid, and the time dependence of load balance efficiency).
Lastly, in \sect{parameter_dependence}, we present the load balancing performance dependence on several algorithm choices and numerical parameters which control aspects of the load balancing.
The simulations presented in this section were run on the Oak Ridge Leadership Computing Facility (OLCF) Summit system (which consists of IBM AC922 server nodes, two IBM Power9 CPUs and six NVIDIA V100 GPUs per node).

\subsection{Laser-Ion Acceleration: Problem Setup}
\label{sec:setup}
Laser-ion acceleration was chosen to study our dynamic load balancing strategy because of its substantial spatial variations of both fields and particle distributions over the dynamic timescales.
Our simulations of laser-ion acceleration are \textit{2D3V}, that is to say the simulation plane ($zx$) is two-dimensional in space, yet we track all three components of the electromagnetic field and particle momenta.
The problem setup we describe here is prototypical for scenarios detailed in Refs. \cite{Obst2017,Huebl2019}.

\begin{figure}[!ht]
  \centering
  \includegraphics[width=\linewidth,trim={0.3cm 0.3cm 0.3cm 0.3cm}]{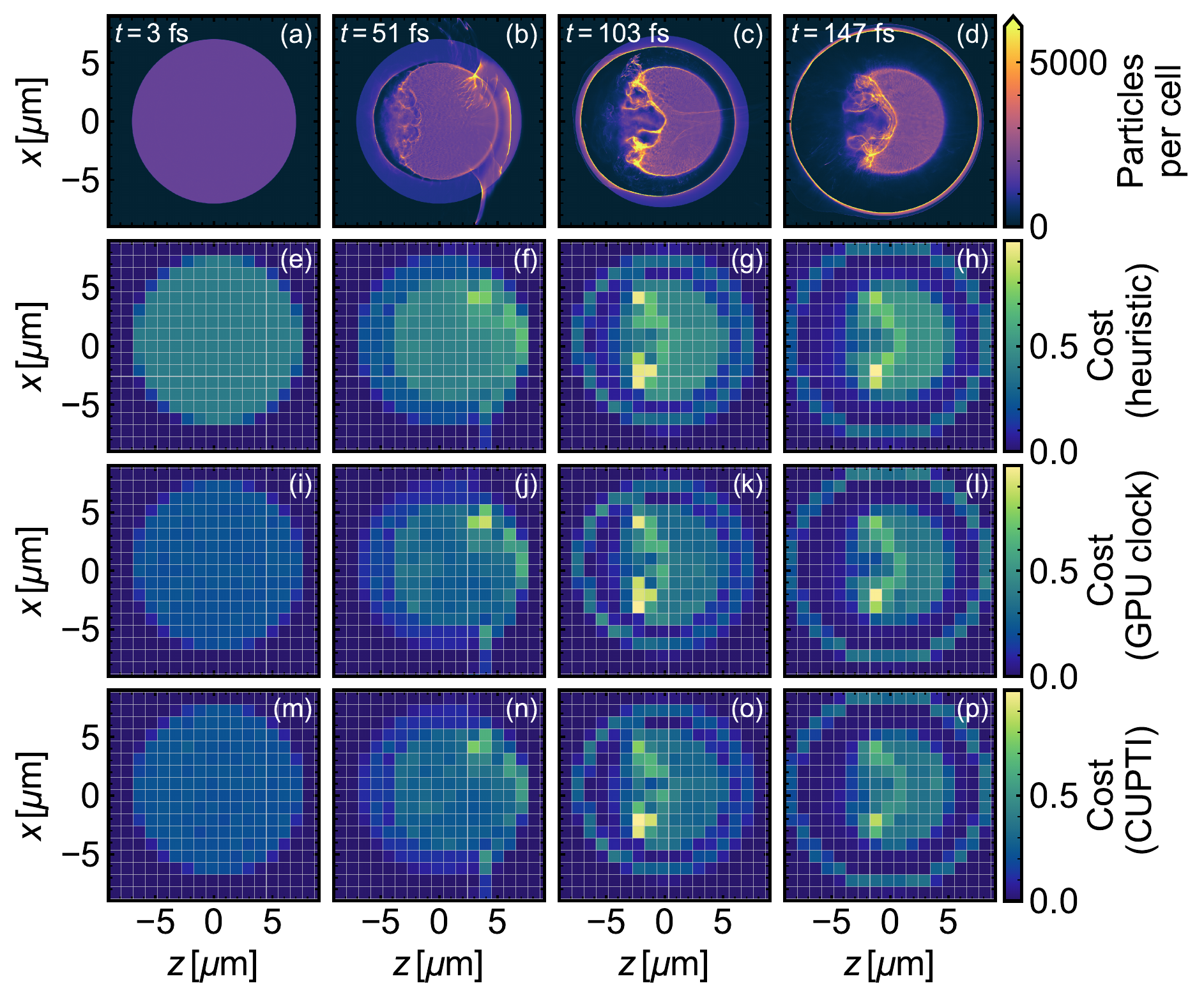}
  \caption{Time evolution of laser-ion acceleration problem (described in \sect{setup}) for number of particles per cell (a)--(d), 
  heuristic cost (e)--(h), 
  GPU clock-measured cost (i)--(l), and 
  CUPTI-measured cost (m)--(p); the computational cost assignment strategies are described in \sect{lb_strategy}. 
  The snapshots are shown at $t=\{3,51,103,147\}\femtoseconds$ for columns 1--4, respectively.
  Gray gridlines in panels (e)--(p) denote the domain decomposition into boxes (see \sect{amrex_overview}).}
  \label{fig:TimeEvolutionAndCosts}
\end{figure}

The initial conditions of our problem are shown in \fig{TimeEvolutionAndCosts}, panel (a). The panel shows a subset of the full computational
 domain, $-15 \micrometers \leq x \leq 15\micrometers$ and $-10\micrometers \leq z \leq 20\micrometers$. 

A dense target of electrons and protons initially fills a circular region of radius $5\micrometers$ (core) + $2\micrometers$ (slope) centered at $x = z = 0\micrometers$. 
The initial number density of electrons and protons in the target is $\ninaut = \nenaut \equiv \nnaut = 8.7 \times 10^{27}{~}\textrm{m}^{-3}$ ($5\times$ the critical plasma density), and cuts off exponentially near the edge of the target over a scale length $L=50\nanometers$. Ions are initialized at rest, whereas electrons are initialized with Gaussian-distributed momenta; the standard deviations of the electron momenta distributions along $x$ and $z$ are both $0.01{~}m_{\rm e} c$ ($\me$ is the electron mass and $c$ is the speed of light). 
An ultrashort laser pulse with a Gaussian profile propagates from $z=-9\micrometers$ along $z$ toward the target starting at $t = 0\seconds$.
Physical parameters of the laser pulse are as follows:  
the peak amplitude of the laser field is $10^{14}{~}\textrm{V}/\textrm{m}$ ($a_0=25$), 
the laser's polarization vector is along $x$, 
the wavelength of the laser pulse is $\lambda_{0} = 800\nanometers$,
the laser beam profile \textit{waist} (i.e., the lateral distance to the laser peak amplitude at which the field amplitude decays by a factor of $e$) is $4\micrometers$, 
the laser pulse duration (i.e., the time required for the laser pulse peak amplitude to decay by a factor of $e$) is $10\femtoseconds$, 
and the laser pulse is initialized over $30\femtoseconds$ and focused at the center of the target.
The simulations (unless otherwise noted) are evolved up to $150\femtoseconds$, which corresponds to around 4100 simulation timesteps; this timescale is long enough to cover the highly kinetic part of the simulated laser-matter interaction physics, and as a result is an appropriate timescale for testing dynamic load balancing.

In addition to the geometry and physical parameters of our setup, detailed above, we specify several numerical parameters.  
The simulation domain $(L_z, L_x) = (30\micrometers, 30\micrometers)$ (unless otherwise noted) is resolved with $(N_z, N_x) = (1920, 1920)$ cells; the cell size is $dz = dx = 0.0195{~}\lambda_{0}$ $= 0.274 \comp$, where $\comp$ is the electron skin depth, and $\omp\equiv\sqrt{\nnaut q_{\rm e}^2 / (\epsilon_{0} m_{\rm e})}$ is the electron plasma frequency; $q_{\rm e}$ is the electron fundamental charge and $\epsilon_0$ is the permittivity of free space.  
For parallel computation of the problem, the domain is decomposed into boxes of size $M_{z} = M_{x} = 64$ cells.
To ensure numerical stability of the finite-difference PIC solver, the time resolution of our simulations is $\Delta t = 0.193\omp^{-1}$, which is less than that required by the Courant-Friedrichs-Lewy condition by a factor of $0.999$.  Lastly, we use $\nppc=900$ macro-particles per cell for each particle species, and third-order particle shapes. With this choice of numerical parameters, the overall time spent in compute-dominated routines is typically about half of the simulation's walltime.
For our fiducial simulation parameters (detailed above), we use 16 nodes (96 GPUs). 
For parallel communications, we use the Message Passing Interface (IBM Spectrum MPI, version 10.3), with one MPI rank per GPU \cite{MPI}.

For completeness, the software, environment, \helvet{WarpX} input files, output data, and analysis tools used to produce and analyze the simulations presented in this paper are archived in Ref. \cite{data}.

\subsection{Load Balance: Time Evolution}
\label{sec:time_evolution}
As the ultrahigh-intensity laser pulse interacts with the dense particle target, electrons respond quickly relative to the more massive hydrogen ions, and penetrate through the solid target. The different response times of electrons and hydrogen ions to the incident laser pulse generates strong electric fields, which can in turn accelerate ions to high energies. 
Throughout this process, a strong kick of the target front by the incident laser results in substantial changes to the spatial density profile as particles are transported through the target; the time evolution for the range $3\mbox{--}147\femtoseconds$ of the spatial profile of particles per cell is shown in \fig{TimeEvolutionAndCosts} (panels (a)--(d)).
The macro-particle number is intentionally kept constant in the lower-density, exponential plasma slope around the target, visible in later steps as a `ring,' for adequate modeling of laser-absorption.
Relative to the initial number of particles per cell in the target ($2\times\nppc=1800$ particles per cell), the density can increase by a factor of $25$ during the simulation (this is true at $t=51\femtoseconds$, shown in panel (b), but note that the color bar range is truncated).

As discussed in \sect{lb_warpx}, particle number density correlates positively with the true compute work (\textit{cost}) associated with a box of the domain. Time variation in the spatial profile of particle number density then implies that the true computational costs, as well as the estimated approximations for each box via the heuristic, GPU clock, and CUPTI proxy schemes (discussed in \sect{lb_warpx}), will change as the simulation progresses. 
The time evolution of the computational cost per box is shown in \fig{TimeEvolutionAndCosts}, rows 2--4, for the three different cost assignment schemes; heuristic (panels (e)--(h)), GPU clock (panels (i)--(l)), and CUPTI (panels (m)--(p)); gray gridlines delimit boxes which comprise the simulation domain. 
Costs along each row have been normalized to the maximum cost per box over the four temporal. 
Comparing the snapshots of costs between schemes confirms they are consistent with one another.  


As discussed in \sect{lb_warpx}, different policies are possible when determining the updated mapping from GPU ownership to boxes; we explore two commonly used policies: knapsack and Morton Z-order space-filling curve (from here on, SFC).
\begin{figure}[!ht]
  \centering
  \includegraphics[width=\linewidth, trim={0.3cm 0.3cm 0.3cm 0.3cm}]{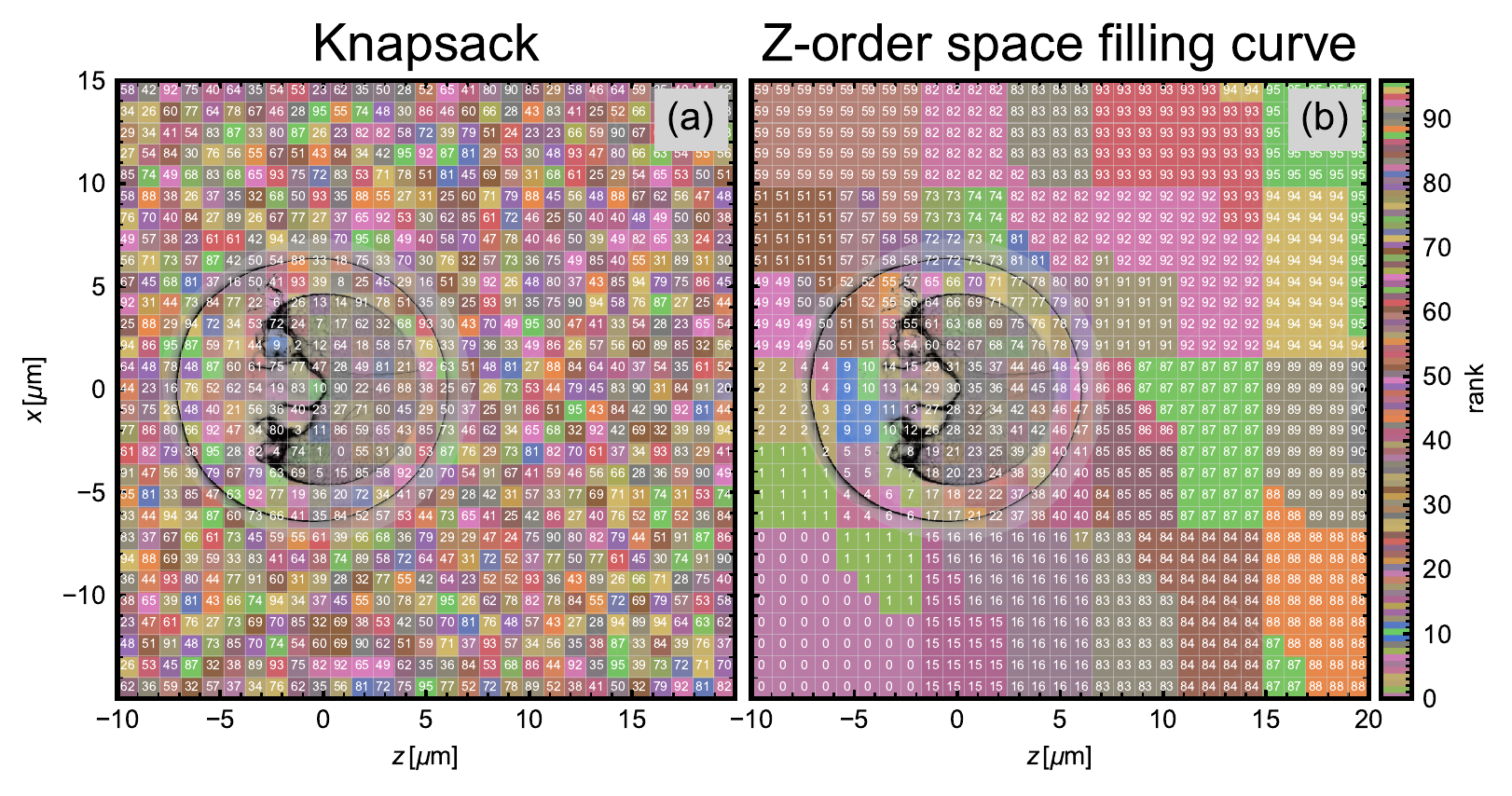}
  \caption{Comparison of distribution mappings that result from knapsack (a) and space-filling curve (b) policies (described in \sect{lb_strategy}), for the laser-ion acceleration problem described in \sect{setup}; a transparent overlay shows the number of particles per cell at physical time $t=103\femtoseconds$ (this corresponds to panel (c) of \fig{TimeEvolutionAndCosts}).}
  \label{fig:DistributionMappings}
\end{figure}
In \fig{DistributionMappings}, we show visualizations of distribution mappings at $t=103\femtoseconds$, computed according to either knapsack or space-filling curve methods (panels (a) and (b), respectively). 96 different colors (shown in the colorbar) correspond to the 96 different GPUs on which this simulation was run (for additional clarity, white GPU ID numbers are printed in the center of each box).
For comparison to the underlying particle distribution, a transparent overlay of the number of particles per cell at $t=103\femtoseconds$ (this corresponds to panel (c) in \fig{TimeEvolutionAndCosts}) has been added to the figure.  

The visualizations of distribution mappings match intuition about the knapsack (panel (a)) and SFC (panel (b)) algorithms. Knapsack calculates its distributed load by grouping boxes as efficiently as possible without any consideration of the spatial location of the boxes; as a result, GPU ownership of boxes appears to be scattered randomly. 
For SFC, boxes are grouped with the constraint that GPU ownership is contiguous along a Z-order curve threading the boxes; as a result, the distribution mapping computed according to the SFC algorithm shows relatively large patches placed on a single GPU in regions with relatively few particles per cell, and smaller patches in more densely packed regions. These small groupings appear roughly in the circular region of radius $7\micrometers$ centered at $(z, x) = (0\micrometers, 0\micrometers)$.
Even though the distribution mappings resulting from the knapsack and SFC policies appear to be strikingly different, the load balance efficiencies (see \eq{lb_efficiency}) attained at this snapshot ($t=103\femtoseconds$) are similar; about $61\%$ and $56\%$ for knapsack and SFC, respectively. With the spatial constraint of the SFC algorithm, the load balance efficiency that is possible with SFC can be no greater than that obtained with knapsack; comparison between SFC and knapsack is further discussed in \sect{parameter_dependence}.

\begin{figure}[ht]
  \centering
  \includegraphics[width=\linewidth, trim={0.35cm 0.3cm 0.3cm 0.3cm}]{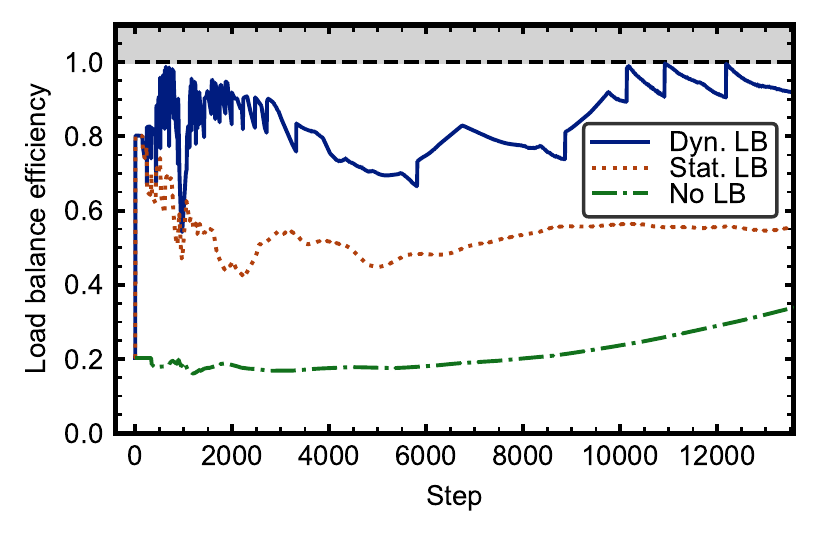}
  \caption{Time evolution of load balance efficiency for simulations with dynamic load balancing (solid blue), with static load balancing (dotted red), and without load balancing (dot-dashed green), for laser-ion acceleration problem similar to that described in \sect{setup} (these simulations run on 4 nodes instead of our fiducial 16 nodes, and are time-evolved up to physical time $t=1000\femtoseconds$, as opposed to our fiducial $150\femtoseconds$). The dynamic and static load balancing runs use knapsack distribution mapping.}
  \label{fig:LoadBalanceEfficiency}
\end{figure}

We show in \fig{LoadBalanceEfficiency} a comparison of the time evolution of load balance efficiency (see \eq{lb_efficiency}) for a simulation with no load balancing (dot-dashed green curve), static load balancing (i.e., the load balancing routine is called once early on in the simulation; dotted red curve), and dynamic load balancing (i.e., the load balancing routine is called periodically as the simulation progresses; solid blue curve); the gray region above the line $y=1$ is not achievable because the load balance efficiency is, by construction, a number on the interval $[0, 1]$ (see \eq{lb_efficiency}).
For the run with dynamic load balancing, the load balancing routine is called once every 10 timesteps (note, as described in \sect{lb_warpx}, this does not necessarily mean that the distribution mapping is updated every 10 timesteps; a proposed distribution mapping is computed at each load balancing step, and is adopted only if it would improve the current load balance efficiency by a prescribed amount. %
In our fiducial simulations and the dynamic load balancing case shown in \fig{LoadBalanceEfficiency}, the proposed distribution is adopted only if the load balance efficiency is at least a $10\%$ improvement relative to the current load balance efficiency).
The distribution mappings for the runs with static and dynamic load balancing are computed according to the knapsack policy.

The simulations shown in \fig{LoadBalanceEfficiency} are similar to the fiducial simulation described in \sect{setup}, but the simulations run for longer physical time (up to $t=1000\femtoseconds$) and have been weak scaled from 16 nodes (96 GPUs) to 4 nodes (24 GPUs). Correspondingly, the grid size is smaller, $(N_{z}, N_{x})=(960, 960)$; at this resolution, the physical duration of the simulation corresponds to about $13600$ timesteps.

It is important to highlight some critical results from \fig{LoadBalanceEfficiency}.  Static load balancing shows a significant improvement in overall load balance efficiency (and correspondingly, the simulation's walltime) compared to the baseline case without load balancing. The average load balance efficiency over the duration of the simulation is $53\%$ with static load balancing and $21\%$ without; static load balancing results in a $2.1{\times}$ speedup, which is similar to the ratio of average load balance efficiencies. 
Dynamic load balancing can yield even greater load balance efficiency throughout the duration of the simulation; with dynamic load balancing, the average load balance efficiency is $84\%$, and yields a $2.9{\times}$ speedup relative to the baseline without load balancing, $1.3{\times}$ speedup relative to static load balancing.

Another critical result from \fig{LoadBalanceEfficiency} is that the dynamic load balancing routine (blue curve), which for this case is called every 10 timesteps, updates the distribution mapping only when it will result in a sufficient improvement to the current load balance efficiency (as described in \sect{parameter_dependence}, our tuned threshold is $10\%$).  This feature can be seen clearly at, e.g., $\textrm{step}=10000$, where the load balance efficiency increases from $91\%$ to $100\%$.
Early on in the simulation ($\textrm{step} \lesssim 3000$), the laser-ion acceleration problem shows rapid changes in the spatial profile of particle number density (and correspondingly, the spatial profile of computational cost), but at later times in the simulation ($\textrm{step} \gtrsim 3000$), the changes occur over longer timescales; these temporal changes are well captured by our dynamic routine.

By requiring that the updated distribution mapping improves the current efficiency by a threshold amount, our implementation avoids the penalty of costly communication of data among ranks (this communication time, when present, dominates the residual time spent in the load balancing routine) when doing so would not substantially improve the load balance.
\subsection{Parameter Dependence of Dynamic Load Balancing Performance}
\label{sec:parameter_dependence}
The performance of typical load balancing implementation depends on several numerical parameters and choices of algorithms, as discussed in \sect{lb_warpx}. Since algorithmic hyperparameters are hard to tune for domain scientists that model a concrete dynamic setup, we investigate multiple approaches both with respect to load balance efficiency, performance as well as minimization of the need for manual user intervention.

\begin{figure*}[ht]
  \centering
  \includegraphics[width=\textwidth, trim={0.3cm, 0.3cm, 0.2cm 0.3cm}]{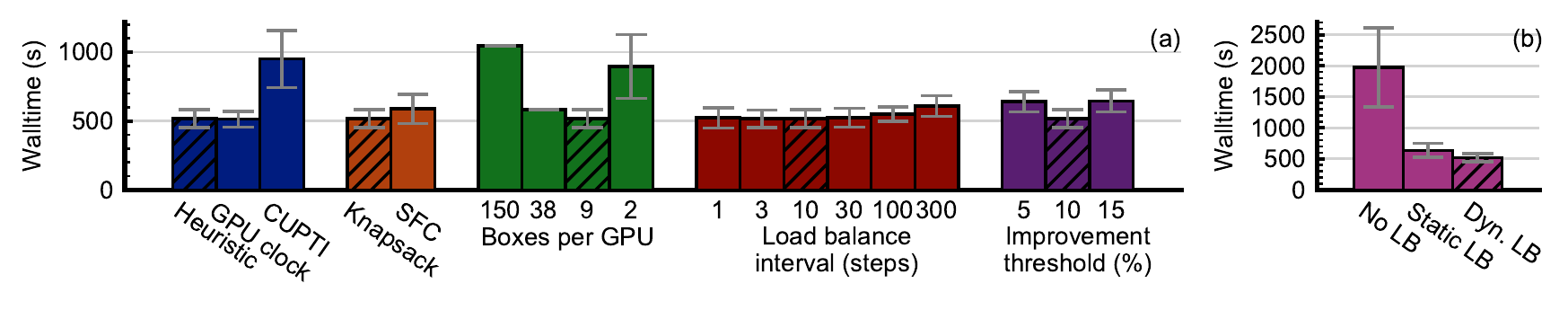}
  \caption{For the laser-ion acceleration problem described in \sect{setup}, parameter and algorithm dependence of walltime (a), and comparison of a case with no load balancing, with static load balancing, and with dynamic load balancing (b).}
  \label{fig:LoadBalanceParameters}
\end{figure*}

In \fig{LoadBalanceParameters}, we present the performance dependence with respect to these parameters and choices of algorithm for our fiducial 16 node simulation; in particular, we show the performance dependence on the choice of computational cost assignment method (heuristic, GPU clock, or CUPTI), the policy used to compute the distribution mapping (knapsack or SFC), the average number of boxes per GPU (150, 38, 9, or 2; for our fiducial domain size $(N_{z}, N_{x})=(1920, 1920)$, this is equivalent to varying the box size $M_{x} = 16, 32, 64,$ or $128$ cells, respectively), the load balance interval (i.e., the inverse of the frequency with which we call the load balancing routine: once every 1, 3, 10, 30, 100, or 300 steps), and the load balance efficiency improvement threshold (i.e., the improvement to load balance efficiency required for a proposed distribution mapping to be communicated and updated: $5\%, 10\%$, or $15\%$).
For each parameter or algorithm scan, we fix those not under examination to the optimal selection from among those shown in panel (a) (for example, for each simulation represented by the first group of bars showing dependence on cost assignment method, we use knapsack, 9 boxes per GPU, load balance interval of 10 steps, and load balance efficiency improvement threshold of $10\%$).

The height of each bar indicates the simulation's walltime in seconds, excluding initialization time ($\approx1$\,s).
Error bars show the spread between minimum and maximum across MPI ranks of the time spent in the most compute-intensive kernel (which includes current deposition and particle push routines); smaller error bars indicate smaller spread between minimum and maximum across MPI ranks of the time spent in current deposition and particle push, and thus correspond to a more balanced load, whereas larger error bars indicate a larger spread, and correspond to a less balanced load.
Note that the hatched bars in \fig{LoadBalanceParameters}  represent the same simulation; this simulation has the highest speedup relative to baseline, with the following hyperparameters: heuristic cost assignment, knapsack distribution mapping, 9 boxes per GPU, load balance interval of 10 steps, and load balance efficiency improvement threshold of $10\%$.

The first group of bars (blue) in panel (a) of \fig{LoadBalanceParameters} shows a comparison between the different cost assignment schemes described in \sect{lb_warpx}, namely heuristic, GPU clock, and CUPTI.
Heuristics need to be tuned by the application user,\footnote{On the OLCF Summit system using finite-difference time-domain field solve and third-order particle shapes, we measured in \helvet{WarpX} particle and cell weights of 0.75 and 0.25, which (unless otherwise stated) we used in the tests presented here. These weights were calibrated based on benchmark tests, one with a relatively large number of particles per cell (27 particles per cell) distributed uniformly over the simulation domain, and another with no particles (only cells); together, these tests yielded estimates of the walltime corresponding to a single particle and cell, and in turn relative weightings for a particle and cell.} while the latter two approaches are free of user-facing hyperparameters for costs.

As discussed in \sect{time_evolution}, the three schemes give similar spatial distributions of computational cost (and as a result, similar measured load balance efficiencies), yet the performance comparison here shows that with the overhead of CUPTI cost collection, the simulation runs about $2{\times}$ slower than with either heuristic or GPU clock cost collection. 
This discrepancy is partly explained by an overhead associated with CUPTI instrumentation;
we find that simply enabling collection of CUPTI activity records and registering a callback function to handle the request and delivery of buffers in which to store activity records (see \sect{lb_warpx}) increases this simulation's walltime by $30\%$; the residual $70\%$ increase in walltime (relative to the heuristic or GPU clock cases) is accounted for mostly by latency in our implementation's data movement of costs to global memory.
Since the CUPTI instrumentation introduces a non-negligible, unmeasured overhead, it is possible that load balancing without knowledge of these unmeasured costs decreases the actualized (as opposed to measured) load balance efficiency.  

The second group of bars (orange) in \fig{LoadBalanceParameters} (panel (a)) shows a comparison between the knapsack and SFC policies (described in \sect{lb_warpx}; see also panels (a) and (b), respectively, of \fig{LoadBalanceEfficiency} for visualizations of sample distribution mappings that result from either policy).
Both the knapsack and SFC simulations presented here employ heuristic cost assignment, but we have tuned the particle and cell weights (see \sect{lb_warpx}) to give the most favorable comparison with knapsack;\footnote{By inspection, we determined (for heuristic cost assignment) that particle and cell weights of 0.02 and 0.98, respectively, yield roughly optimal performance with the SFC algorithm; we note that the \helvet{AMReX} implementation of the knapsack algorithm includes the option to limit the maximum number of boxes per GPU (our default is 1.5 times the average number of boxes per GPU), but this option is not implemented for SFC.
The relatively large cell weight and small particle weight that we find are optimal with SFC emulates the constraint on the maximum number of boxes per GPU.}
at best, we find that the walltime with SFC is about comparable to that obtained with the knapsack algorithm.
The lesser walltime obtained with knapsack, relative to that obtained with SFC, is perhaps counterintuitive; by construction, the SFC algorithm increases the correlation length with respect to GPU ownership of boxes over the simulation domain (as in panel (b) of \fig{LoadBalanceEfficiency}, note the relatively large unicolored patches), and thus reduces overall costs associated with communications; while the geometric constraint of the SFC algorithm may result in lower load balance efficiency relative to knapsack, the effect of improved communications with SFC may outweigh the incurred penalty to load balance efficiency, leading to a net reduction in simulation walltime.
This scenario requires that communications account for a significant fraction of the simulation's walltime, which is indeed true in our case: as discussed in \sect{setup}, compute-intensive kernels account for $50\%$ of the measured walltime, and the remainder is predominantly accounted for by communication routines. 
Still, relative to knapsack, we see no speedup with SFC.  

To partially account for this, we note that in \helvet{AMReX}'s communication routines, a patch of boxes posts non-blocking MPI send operations to move data that must be communicated to a different GPU, then performs local work (such as local copy operations required by interior boxes), and lastly blocks until all MPI requests have completed.  
Although SFC  has the potential to improve intra-GPU communication, the overall time is still bound by the inter-GPU data transfer required by boxes at patch edges.
This may partially account for the comparable performance of knapsack and SFC. 

The third group of bars (green) in \fig{LoadBalanceParameters} (panel (a)) shows the effect of varying the average number of boxes per GPU\footnote{Note that the number of boxes per GPU, as opposed to the \textit{average} number of boxes per GPU, is not fixed;  this can of course vary as GPU ownership of boxes is shuffled due to load balancing.} (for a fixed domain size, this is equivalent to varying box size), which we vary over 150, 38, 9, and 2 (these correspond, respectively to box sizes $M_{x} = 16, 32, 64,$ or $128$ cells). 
Increasing the average number of boxes per GPU (decreasing the box size) produces a trade-off between the overhead associated with managing more boxes (for example, decreasing the box size on a fixed domain increases the total number of guard cells and communication becomes more costly), and the improved load balancing that is possible with a smaller box size (note that smaller boxes enable more fine-grained pixelization of the spatial profile of cost, and thus greater load balance efficiency relative to larger boxes); the improved load balancing can be inferred from the shrinking error bars with decreasing box size (equivalently, increasing average number of boxes per GPU).
Even though the load balance efficiency is greater with an average of, e.g., 150 boxes (of size $M_{x}=16$) than with 2 boxes (of size $M_{x}=128$ cells) per GPU, the walltime is greater ($896\seconds$ with 150 boxes per GPU as opposed to $1040\seconds$ with 2 boxes per GPU).  This is because the overhead associated with a greater number of boxes (of smaller size) outweighs the performance benefit of improved load balance efficiency.
We find that an average of 9 boxes per GPU (box size $M_{x}=64$ cells) produces a close to optimal trade-off between overhead and improved load balancing capability.

The fourth group of bars (red) in panel (a) of \fig{LoadBalanceParameters} shows the performance dependence on load balance interval, which we vary over 1, 3, 10, 30, 100, and 300 steps (load balance interval of 10, for example, means the load balancing routine is called once every 10 steps).  
As discussed in \sect{time_evolution}, there is little penalty to walltime when calling the load balancing routine frequently because the costly operation of communicating and updating the distribution mapping on all ranks is done only when doing so would improve the load balance efficiency more than a minimal threshold.

This improvement threshold is shown in the last group of bars (purple) in panel (a) of \fig{LoadBalanceParameters}, which we vary over $5\%, 10\%$, and $15\%$. If the improvement threshold is too low, communications become more costly as the distribution mapping is updated more frequently; on the other hand, if the improvement threshold is too high, the distribution mapping is not updated frequently enough to reap the performance benefit of load balancing.  We find that an improvement threshold of $10\%$ offers an optimal trade-off between these competing effects.

For the simulations we present here, the time required to gather costs from all ranks (which is needed to determine whether to update the distribution mapping) is, at most, no greater than $2.3\%$ of the total walltime (which is achieved with a load balance interval of 1).
Since this is a relatively small fraction of the total walltime, the load balancing routine may be called frequently to ensure that load balance is maintained, without incurring a significant penalty to walltime.
Empirically, we find little difference in walltime for load balance intervals in the range 1--30 steps, with an increasing trend for load balance intervals $\gtrsim 30$ steps (this is due to the lower load balance efficiency on average over the duration of the simulation). 

To summarize, we achieve best performance with the following selection of parameters and algorithms: GPU clock cost assignment, knapsack policy for computing the distribution mapping, an average of 9 boxes per GPU (equivalently, a box size $M_{x}=64$ cells), and calling the load balancing routine once every 10 steps (similar performance is achieved with heuristic in place of GPU clock cost measurement, however this requires the user to tune particle and cell weights, which can vary depending on hardware an algorithmic choices). 
With these selections, load balancing accounts for (at most, across all ranks) $4$--$6\%$ of the walltime in our test problem.

In panel (b) of \fig{LoadBalanceParameters}, we compare the walltimes of a baseline case without load balancing, a case with static load balancing (i.e., the load is balanced once toward the beginning of the simulation), and a case with dynamic load balancing where the parameters and algorithms are tuned for roughly optimal performance (as determined from the tests shown in panel (a); the static load balancing case has the same parameter and algorithm selections as the dynamic load balancing case, apart from the load balance interval).  
The relatively large error bars for the baseline case without load balancing indicate a severe load imbalance.
Static load balancing results in a $2.9{\times}$ speedup relative to the baseline case without load balancing;
dynamic load balancing yields even greater performance improvement, with a $3.8{\times}$ speedup relative to the baseline case without load balancing ($1.2{\times}$ speedup compared to static load balancing). 
This shows that the spatial profile of costs varies substantially with time, and that dynamic, as opposed to static, load balancing is essential to improved performance.


\section{Performance Assessment and Scaling of Load Balancing}
\label{sec:performance}
In this section, we present a performance model, calibrated with strong scaling measurements of our code, and use it to assess the performance of our improvements to \helvet{WarpX}'s load balancing in the laser-ion acceleration problem described in \sect{setup}.
We present the weak scaling of our load balancing routine from 1 up to 1024 nodes (6 -- 6144 GPUs).  As in \sect{test_problem}, these simulations were run on the OLCF Summit system (IBM AC922 server nodes; 2 IBM Power9 CPUs and 6 NVIDIA V100 GPUs per node).  

For a given code, how much of a performance improvement may be anticipated with load balancing?  
To answer this question, we consider the load balancing operation in the context of strong scaling, i.e. increasing compute resources for a problem of fixed size.
With measurements of strong scaling for a given code, one may model the performance response as a function of available compute resources; ideally, performance improves linearly with increasing compute resources, and with a constant of proportionality equal to unity. 
However, in realistic codes, factors such as communications and overhead may modify the relationship and lead to less than ideal scaling.
A strong scaling model can capture these nonideal effects and be used to predict the performance improvement that would result from a given increase in compute resources.

To apply strong scaling in understanding the performance improvement that is possible through load balancing, we consider a simulation with a computational load that is initially imbalanced (i.e., there is at least one compute element assigned a greater computational load than another); we call this initial maximum compute work (over all compute elements) $c_{\rm max0}$. 
The performance (e.g., walltime) of this hypothetical simulation is limited by the initial load imbalance $c_{\rm max0}$, and may be improved by distributing compute work evenly over available resources.  
After perfect load balancing, the compute work is equally distributed over all $N$ compute elements, $c_0 = c_1 = \cdots = c_{N} = c_{\rm avg0}$, which becomes the new performance limiter (we define $c_{\rm avg0}$ as the average at initialization of $c_0, c_1, \cdots, c_{N}$).
Load balancing is similar to strong scaling: more compute resources are assigned to the work $c_{\rm max0}$, and the workload of the compute element initially assigned that work decreases from $c_{\rm max0}$ to $c_{\rm avg0}$; with strong scaling, increasing the compute resources while holding the problem size fixed effectively decreases the compute work assigned to each compute element.

\begin{figure}[!ht]
  \centering
  \includegraphics[width=\linewidth, trim={0.35cm 0.3cm 0.cm 0.2cm}]{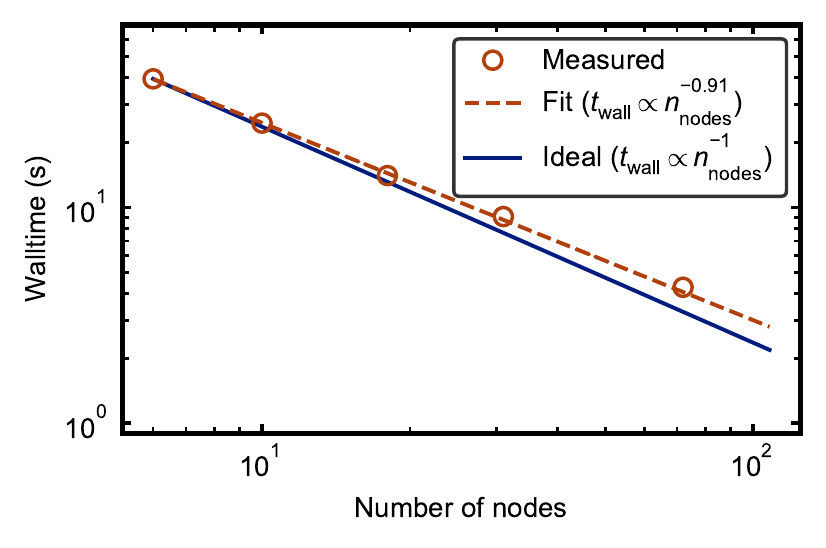}
  \caption{Strong scaling for a \helvet{WarpX} baseline problem (domain filled uniformly with plasma).
  Circles show measurements and the solid curve shows ideal scaling of walltime with number of compute nodes. 
  The dashed curve shows a fit to the measured walltimes, which we use, along with the level of initial load imbalance, to predict the maximum possible speedup possible with perfect load balancing.}
  \label{fig:StrongScaling}
\end{figure}

The ratio $c_{\rm max0}/c_{\rm avg0}$ can therefore be calibrated (weighted) against a code's characteristic strong-scaling efficiency, which we fit to an exponential model for walltime: $t_\mathrm{wall}\approx n^{-x}_\mathrm{nodes}$.
The parameter $x$ is a value between $1$ (ideal) and $0$.
The maximum speedup, $S$, from perfect load balancing can then be expressed as: 
\begin{align}
    S &= \left(\frac{c_{\rm max0}}{c_{\rm avg0}}\right)^{x} 
    = \left(\frac{1}{E_{0}}\right)^{x},
\end{align}
where $E_{0}$ is the initial load balance efficiency of the simulation (see \eq{lb_efficiency}).
For example, for $16$ nodes, we measure load a imbalance ratio to average cost of $c_{\rm max0}/c_{\rm avg0} = 6.2$.
Perfectly load-balanced on the same amount of compute resources, this is the factor the slowest compute element is strong-scaled.
Calibrated against \helvet{WarpX} characteristic strong scaling, in 2D3V $x=0.91$ and 3D3V $x=0.88$, the best speedup a load balance algorithm can achieve with this starting condition and assuming a dynamically sustained imbalance is $5\times$.
In practice, realization of this maximum speedup may be limited by further factors, such as the cost of performing a load balance operation and more complex communication patterns.\footnote{In our case, decreasing the box size $M_{x}$, as discussed in \sect{parameter_dependence}, leads to an overall performance penalty, in spite of the improved load balancing with smaller box size (see the third group of bars (green) in panel (a) of \fig{LoadBalanceParameters}).}

In \fig{StrongScaling}, we show strong scaling measurements (red circles) for \helvet{WarpX}, i.e., walltime as a function of compute nodes; the problem setup is a domain of size $(N_{z}, N_{x}) = (3072, 3072)$ cells, filled uniformly with 550 particles per cell.
In the tests, we vary the number of nodes over 6, 10, 18, 31, and 72 (36, 60, 108, 186, and 432 GPUs, respectively).  
The dashed line shows a log-log fit to our measurements, yielding a performance model for walltime, $t_{\rm wall} \propto n_{\rm nodes}^{-0.91}$ (for comparison with the ideal strong scaling, we show also the curve $t_{\rm wall} \propto n_{\rm nodes}^{-1}$ as a solid line).

\begin{figure}[!ht]
  \centering
  \includegraphics[width=3.25in, trim={0.3cm 0.3cm 0.2cm 0.2cm}]{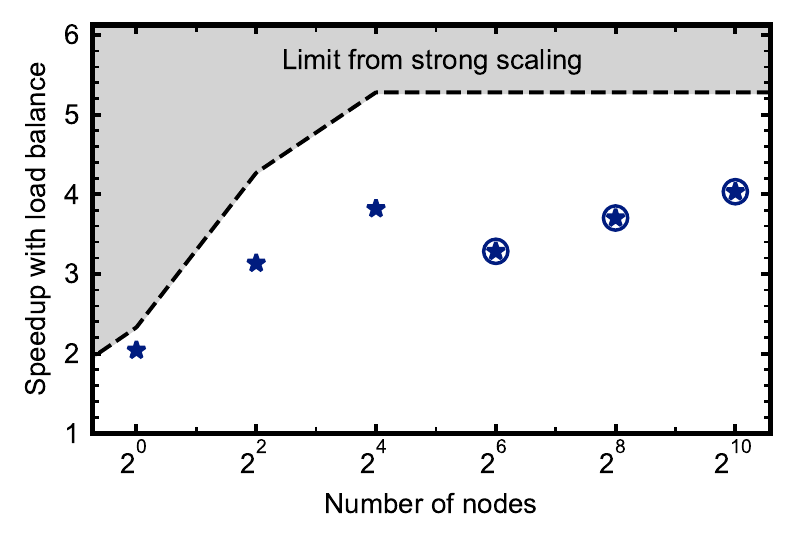}
  \caption{For the laser-ion acceleration test problem described in \sect{setup}, weak scaling (from 1 up to 1024 nodes, i.e., 6 up to 6144 GPUs) of speedup factor, relative to baseline case without load balancing.
  The dashed line shows maximum possible speedup, as predicted from strong scaling and the initial level of load imbalance. 
  Circled points indicate cases in which the no-load balancing baseline exceeded GPU memory prior to completion of the simulation ($43\%, 9\%,$ and $8\%$ completion for $64, 256$, and $1024$ nodes, respectively); for these cases, the simulations with dynamic load balancing ran to completion without exceeding GPU memory capacity.}
  \label{fig:WeakScaling}
\end{figure}

To assess the performance of \helvet{WarpX}'s updated load balancing, we performed a weak scaling test (from 1 up to 1024 nodes, i.e., 6 up to 6144 GPUs) of the laser-ion acceleration problem with and without load balancing (described in \sect{setup}), and compared the measured speedup (i.e., the ratio of walltime with load balancing to walltime without load balancing; blue points in \fig{WeakScaling}) to the ideal limit computed from strong scaling and the initial load imbalance (dashed line in \fig{WeakScaling}).
For $64, 256$, and $1024$ nodes, our simulations without load balancing exceeded a GPU's memory (16 GB on the NVIDIA V100 GPUs we use) before reaching the prescribed final timestep, leading to extremely degraded performance (this is indicated by the circled points in \fig{WeakScaling});

for these cases the speedup is computed only with walltimes (of the load balanced and load imbalanced cases) up to the timestep that the load imbalanced case exceeds a GPU's memory.\footnote{We also ran the $64$ and $256$ node cases (both with and without load balancing) with fewer particles per cell, such that the cases without load balance ran to completion without exceeding GPU memory; the measured speedup factors are similar to those shown by the circled points in \fig{WeakScaling}, which indicates that those measurements, though limited to the period before exceeding GPU memory, are representative of the speedup possible over the full prescribed time range.}
For $64, 256$, and $1024$ nodes, the simulations without load balancing reached $43\%, 9\%,$ and $8\%$ completion, respectively, before exceeding a GPU's memory; this demonstrates that, especially when a code uses memory-limited GPUs, load balancing is not just crucial to performance: it enables simulations that are not possible otherwise for a given amount of compute resources!
With static load balancing, we tested the cases with $64$ and $256$ nodes and they ran to completion without exceeding GPU memory capacity, but for problems which develop severe load imbalance as the simulation progresses, dynamic load balancing may be necessary to prevent the performance-breaking effect of exceeding GPU memory capacity.

For number of nodes in the range $4$--$1024$, we see, relative to the baseline case without load balancing, a $3\textrm{--}4{\times}$ speedup with dynamic load balancing (which is likely an underestimate for the crashing runs with $64$, $256$, and $1024$ nodes); compared to the ideal maximum speedup computed from strong scaling, these measurements attain from $62\%$ to $74\%$ of the predicted limit (when running on 6 GPUs, we observe a $2{\times}$ speedup, which corresponds to $88\%$ of the predicted maximum speedup).
%


This study has demonstrated that adaptive, run time based GPU timers yield a substantial performance gain for \helvet{WarpX} PIC simulations with temporally variable particle distributions. GPU-based simulations of next generation particle accelerators and astrophysical plasmas can now readily take advantage of these features to ensure an efficient and performant transition to anticipated exascale supercomputers.

\section{Related Work}
\label{sec:related}

State-of-the-art particle-in-cell codes \cite{OSIRIS2002, Spitkovsky2005, VPIC2008, PIConGPU2013, Smilei2018, Miller2020} typically support either static load balancing or dynamic load balancing targeting CPUs.
To the knowledge of the authors, those particle-mesh codes that do support dynamic load balancing on GPUs rely on cost functions based on node-local application data as a surrogate for computing run time \cite{Tsuzuki2016, Germaschewski2016}.
The heuristic approach, however, is not necessarily an accurate reflection of true computing run time \cite{SAMRAI2001}.
To gain the full performance benefit of dynamic load balancing, without excessive tuning of hyperparameters by the user, computational costs can be measured accurately at runtime.
An adoption of dynamic load balancing strategies for accelerator hardware in exascale-workflows is therefore needed, including a reduction of user-facing hyperparameters, which are hard and expensive to tune in realistic scenarios.

Dietrich et al. \cite{Dietrich2012} demonstrated instrumentation of CUDA kernels as a way to measure kernel run times for post-run analysis.
One of the herein presented implementations for cost assessment which is free of user-facing hyperparameters, GPU clock, employs a similar approach, yet provides measurement of costs in-situ as the application is running, thus enabling on-the-fly load balancing based on measured computational costs (see \sect{lb_warpx}).
Furthermore, the runtime overhead introduced with this technique is small enough that it does not negate the performance benefit of automatic runtime load balancing \cite{Wijngaart2017}.
The GPU clock strategy for cost measurement is potentially portable, given the accelerator framework supports a clock method (or similar) to assess kernel run times.

\section{Summary and Discussion}
\label{sec:conclusion}
In this work, we present enhancements to dynamic load balancing for particle and mesh-based simulations targeting GPU architectures.
As a component of these improvements, we introduce several GPU-applicable strategies for measuring the relative computational costs of sub-domains of compute work.
While as application developers we can find optimal hyperparameters for a specific setup for heuristic cost-functions, we demonstrate that a measurement of the actual kernel run time can be established.
Especially, we implemented a potentially vendor-neutral, in-situ, in-kernel measurement of run time based on a GPU clock that shows negligible load-balancing overhead in practice.
Contrarily, a measurement approach based on Nvidia CUPTI added significant run time overhead and is not vendor-neutral.

We demonstrate our methods in the fully kinetic particle-in-cell code \helvet{WarpX} and explore its performance.
For the scientifically relevant test case of laser-ion acceleration, we explore the performance dependence of load balancing on several numerical parameters and algorithm choices that enter into our routine, including 
cost assignment strategy (heuristic cost assessment based on a weighted linear sum of the number of particles and cells, GPU clock timing to assess summed thread execution time, and CUPTI-based timing to assess kernel execution time using NVIDIA's CUPTI API), load balance strategy (knapsack or SFC), 
the number of boxes per GPU (equivalently, on a fixed domain, the box size in cells), the frequency with which we call the load balancing routine, and the load balance efficiency improvement required to communicate and update a proposed distribution mapping. 
For the laser-ion acceleration that is the focus of the present work, we measure a $3.8{\times}$ speedup relative to the baseline without dynamic load balancing and a $1.2{\times}$ speedup compared to the static load balancing baseline. 

To assess the performance of \helvet{WarpX}'s updated dynamic load balancing, we introduce a performance model based on strong scaling measurements of \helvet{WarpX} and link performance improvement through strong scaling to the initial level of load imbalance in our simulations, thereby predicting a theoretical maximum speedup factor that is achievable through load balancing. 
We present the weak scaling (from 24 up to 6144 GPUs on Summit) of our dynamic load balancing performance relative to the baseline case without load balancing, and find that dynamic load balancing improves performance, with achieved improvement typically $62\%$ to $74\%$ of the predicted maximum ($88\%$ when running on 6 GPUs); in particular, several of our simulations demonstrate that dynamic load balancing is de-facto a prerequisite for productive usage of distributed, locally limited GPU memory at scale.

In the present work, we focused on load balancing according to the cost of GPU compute.
Incorporating communication costs into our load balancing routine will be a topic of future investigation, and may bring the performance of \helvet{WarpX}'s load balancing closer to the theoretical limit of our presented performance model.
Further exploration of communication costs may also help to demystify the relative performance of knapsack and SFC update for the distribution mapping, which remains an open question. 

\begin{acks}
The authors thank the \helvet{WarpX} and \helvet{AMReX} development teams for invaluable contributions.
We thank Andrew T. Myers and Weiqun Zhang for valuable discussions.
An award of computer time was provided by the ASCR Leadership Computing Challenge (ALCC) program. 
This research used resources of the Oak Ridge Leadership Computing Facility, which is a DOE Office of Science User Facility supported under Contract DE-AC05-00OR22725.
This research also used resources of the National Energy Research Scientific Computing Center (NERSC), which is supported by the Office of Science of the U.S. Department of Energy under Contract No. DE-AC02-05CH11231.
This research was supported by the Exascale Computing Project (17-SC-20-SC), a joint project of the U.S. Department of Energy's Office of Science and National Nuclear Security Administration, responsible for delivering a capable exascale ecosystem, including software, applications, and hardware technology, to support the nation's exascale computing imperative.
This work was performed in part under the auspices of the U.S. Department of Energy by Lawrence Berkeley National Laboratory under Contract DE-AC02-05CH11231.
\end{acks}

\bibliographystyle{ACM-Reference-Format}
\bibliography{references}

\end{document}